\documentclass[twocolumn,showpacs,preprintnumbers,amsmath,amssymb]{revtex4}
\usepackage{graphicx}
\usepackage{dcolumn}
\usepackage{bm}

\begin{document}


\title{Representation Mixing and Exotic  Baryons  in the Skyrme Model}

\author{Hyun Kyu Lee }
  \email{hyunkyu@hanyang.ac.kr}
\author{Ha Young Park}
    \email{hayoung@ihanyang.ac.kr}
\affiliation{Department of Physics , Hanyang University, Seoul
133-791, Korea.}

\date{\today}

\begin{abstract}
We study the effect of representation mixing in the $SU(3)$ Skyrme
model by diagonalizing exactly the representation-dependent part.
It is observed that even without the next-to-leading order
symmetry breaking terms  the low-lying baryon masses as well as
the recently discovered $\Theta^+$ and $\Xi_{\bar{10}}$  can be
fairly well reproduced within $3\%$ accuracy.  It is also
demonstrated that the mixing effect is not negligible in decay
processes of $\{\bar{10}\}$. In particular the effect of mixing
with $\{ 27 \}$ is found to be quite large. These results are
compatible with the second-order perturbation scheme. The decay
widths are found to be sensitive to the  mass values. The decay
widths of $\{\bar{10}\}$ are estimated to be smaller than those of
$\{10\}$ by an order of magnitude  due to the destructive
interference between operators although the kinematic factors are
comparable.
\end{abstract}

\maketitle

The recent discoveries\cite{exp} of $\Theta^+$  have generated
lots of interesting  developments in hadron spectroscopy, in
particular in understanding the exotic nature of the state. The
state is exotic in the sense that the quantum numbers   cannot be
explained as a system of three quarks, as the smallest number of
quarks consistent with $\Theta^+$ is five, or that it cannot be
classified into conventional classifications, $\{ 8\}$ and $\{ 10
\}$. The lowest multiplets consistent with $\Theta^+$ is $\{
\bar{10} \}$ in the scheme of flavor $SU(3)$ symmetry.

The chiral soliton model proposed by Skyrme\cite{skyrme} has been
explored theoretically and phenomenologically with many
interesting
successes\cite{witten}\cite{ANE}\cite{guadagnini}\cite{p0} in
describing the properties of low-lying hadrons.   The importance
of the higher multiplets beyond octet and decuplet has been
noticed in the chiral soliton model in treating the symmetry
breaking part as perturbations.  The symmetry breaking part is not
diagonal in the
$SU(3)$ multiplet space so that in higher order
perturbation\cite{p0}\cite{praszalowicz1}\cite{PJH} or
in diagonalizing the hamiltonian\cite{an}\cite{kll} the mass
eigenstate should be mixed with higher representations. For
example, the nucleon  is dominantly described by $\{8\}$ but with
non-vanishing mixing amplitudes of $\{ \bar{10} \}, \{ 27 \},
\cdots $ and the $\Delta$ also  has non-vanishing mixing
amplitudes of $\{ 27 \}, \{ 35 \}, \cdots $.

The prediction\cite{praszalowicz1}\cite{dpp} of $\Theta^+$ as the
lowest state  among the higher multiplet  $\{ \bar{10} \}$ has now
been confirmed. One of the characteristics of $\Theta^+$  as an
isospin singlet and hypercharge 2 state with respect to
representation mixing is that  it has no corresponding state in
the $\{8\}$ and $\{ 27 \}$, {\it i.e.},   no representation mixing
is possible. On the other hand more massive states in the same
mulitiplet $\{ \bar{10} \}$ have non-negligible  mixing with other
representations and the masses and decay widths are supposed to
depend on the mixing.  The effect of mixing in second order
perturbation has been extensively discussed recently
\cite{ekp}\cite{praszalowicz2}\cite{bb}, in which the effect of
mixing is found to be  non-negligible but depends much on the
parameters of the underlying  effective theory. The similar
observations have been made in the exact diagonalization
method\cite{Wallisev}\cite{WK} for the exotic baryon masses. In
this short note we discuss the mixing effect on the decay process
further using the exact diagonalization method keeping only the
chiral symmetry breaking term that is of leading-order in $N_c$.

The effective action for the pseudoscalar mesons, which realizes
the global $SU(3)_L \times SU(3)_R$ in the Goldstone mode, can be
written in general as
\begin{eqnarray}
{\cal S}_{eff} &=& {\cal S}_2 + {\cal S}_{HOD}  + {\cal S}_{SB} +
{\cal S}_{WZ}
\end{eqnarray}
where ${\cal S}_2 $ and ${\cal S}_{HOD}$ are the leading kinetic
term and the higher order derivative terms including the Skyrme
term. ${\cal S}_{WZ}$ is the Wess-Zumino action and  ${\cal
S}_{SB}$ is an explicit symmetry breaking term depending on the
meson masses. The effective Hamiltonian  after quantizing the
``degenerate rotational mode" of the $SU(2)$ soliton of hedge-hog
ansatz\cite{skyrme} embedded into $SU(3)$ is known to have the
following form for $N_c =3$ and $B=1$:
\begin{eqnarray}
H=&&M_{cl}+\frac{1}{2}(\frac{1}{I_1}-\frac{1}{I_2})C_2(SU(2)_R)-\frac{3}{8I_2}\nonumber\\
  &&+ \frac{1}{2I_2}C_2(SU(3)_L) - \alpha(1 - D^8_{88}(A))\label{H},
\end{eqnarray}
where $C_2(SU(2)_R)$ and $C_2({SU(3)}_L)$ are the corresponding
Casimir operators $( C_2(SU(2)_R) = J(J+1), ~~ C_2({SU(3)}_L) =
\frac{1}{3}[p^2 + q^2 + 3(p+q)+ pq]$. In this frame work we are
left with four parameters, $M_{cl}, I_1, I_2$ and $\alpha$, which
should be in principle determined unambiguously from the effective
action.  In this work however we take them as a set of free
parameters for the phenomenological study.

In eq.(\ref{H}), the $SU(3)$ symmetric limit can be achieved when
the last term vanishes.  The mass spectrum of the baryon can be
determined by treating the  symmetry breaking term in a
perturbative way.  One can also include additional  terms
next-to-leading order in chiral symmetry breaking to reproduce the
low-lying baryon spectrum well in the first order perturbation
calculation\cite{dpp}. In this work when the hamiltonian is to be
diagonalized, we do not include these terms which are of next-to
leading order in the $1/N_c$ expansion to make the analysis free
from possible  ambiguities due to  the extra parameters in the
effective theory. For the diagonalization the hamiltonian can be
divided into two parts, representation independent($H_0$)  and
dependent($H_R$) parts: \begin{eqnarray} H_0 &=& M_{cl} +
\frac{1}{2} (\frac{1}{I_1} - \frac{1}{I_2})
C_2(SU(2)_R) - \frac{3}{8I_2} \\
H_R &=& \frac{1}{2I_2}C_2(SU(3)_L) - \alpha(1 - D^8_{88}(A)).
\end{eqnarray} Minimal extensions beyond octet and decuplet can be guided
by considering the quark content of the baryons.   Three quark
system leads  up to decuplet.  With an additional quark-antiqauark
pair for a penta-quark system, $qqq~ \bar{q}q$,  the possible
representation can be extended up to  $\{ \bar{10} \}, \{ 27
\}$,and $ \{ 35 \}$.  With the constraint $Y_R =1$ for $B=1$
baryon, the
state vectors\cite{kll}
for spin
$1/2$ baryons and spin$3/2$ baryons can be written as
\begin{eqnarray}
|B(J = \frac{1}{2}) \rangle = C^a_8 |\Psi ^8 _{\mu , \nu} \rangle
+ C^a_{\bar{10}}|\Psi^{\bar{10}}_{\mu, \nu}\rangle
+ C^a_{27}|\Psi^{27}_{\mu , \nu} \rangle, \label{mix1}\\
|B(J = \frac{3}{2}) \rangle = C^b_{10} |\Psi ^{10} _{\mu , \nu}
\rangle + C^b_{27}|\Psi^{27}_{\mu, \nu}\rangle +
C^b_{35}|\Psi^{35}_{\mu , \nu} \rangle,\label{mix2}
\end{eqnarray}
where a(b)  refer to a baryon with flavor part $\mu = ( Y, I,
I_3)$ and spin part $\nu = (Y_R , J, J_3)$  with spin $J =
1/2(3/2)$. By diagonalizing the hamiltonian, $H_R$,  in the form
of $3\times 3$ matrix for each baryon state, we can calculate the
corresponding mass as an eigenvalue of the hamiltonian.

The eigenvalue and the mixing amplitudes in eqs.(\ref{mix1}) and
(\ref{mix2}) are of course functions of four parameters, $M_{cl},
I_1, I_2$ and $\alpha$. We fix the parameters by a best fit to the
masses of the low-lying octet and decuplet states. The best fit to
the mass differences can be obtained with the central value of
$I_2 = 2.91 \times 10^{-3}$ MeV$^{-1}$and $\alpha = -750$ MeV.
Then the mass fit gives $M_{cl}= 773$ MeV and $I_1=6.32 \times
10^{-3}$MeV$^{-1}$.  It is interesting to note that these values
are comparable to those used in the perturbation
scheme\cite{ekp}\cite{praszalowicz2}.

The masses in the best fit are given by
\begin{eqnarray}
M(N)~~=~939, &M(\Lambda)~~= 1108, &M(\Sigma)~~=1226, \nonumber \\
M(\Xi)~~=1345, &M(\Delta)~~= 1231, &M(\Sigma_{10})=1385,\nonumber\\
M(\Xi_{10})= 1506, &M(\Omega)~~= 1638,
&M(\Theta^{+})=1570,\nonumber\\ M(N_{\bar{10}})= 1705,
&M(\Sigma_{\bar{10}})=1811, &M(\Xi_{\bar{10}})=1818. \label{mass}
\end{eqnarray}
One can see that the masses for the low-lying octet and decuplet
are reasonably well reproduced in the exact diagonalization
method, with results that are comparable also to those obtained in
the perturbation scheme(either in the first order\cite{dp} or in
the second order perturbation\cite{ekp}). It is found that the
estimated masses of $\Theta^+$ and $\Xi_{\bar{10}}$ are consistent
with the experimental values within $3\%$ accuracy.  The mixing
amplitudes for the corresponding states can be read out from the
normalized mass eigenstates.  For example, the mixing amplitudes
for $N, \Delta$ and $N_{\bar{10}}$ are given by
\begin{eqnarray}
C^{N}_{8} = 0.953 ,& C^{N}_{\bar{10}} = 0.234, & C^{N}_{27} =
0.191, \nonumber  \\
C^{\Delta}_{10} = 0.877, & C^{\Delta}_{27} = 0.464 , &
C^{\Delta}_{35}
= 0.125, \nonumber \\
C^{N_{\bar{10}}}_{8} = -0.234 ,& C^{N_{\bar{10}}}_{\bar{10}} =
0.970 ,
  & C^{N_{\bar{10}}}_{27} =
0.024, \label{mixinga}
\end{eqnarray}
which are comparable to those in \cite{ekp}\cite{dp}\cite{bb}. For
$\{ \bar{10} \}$, it should be noted that the equal spacing rule
in the first order perturbation is not literally respected due to
the effects of the mixing in the 2nd order perturbation\cite{ekp}.
It is observed that there are no appreciable differences  in  the
mixing amplitudes between the exact diagonalization scheme and 2nd
order perturbation scheme,which is consistent with the high order
perturbative calculations\cite{PJH}.

Given the wave function in the representation space,
eqs.(\ref{mix1}) and (\ref{mix2}), the decay width of a  baryon B
into a low-lying $B'$ and meson $\varphi$ can be obtained by
evaluating the matrix element of the baryon decay operators. The
Yukawa coupling in general as well as  the decay operator in
particular, which is basically a meson-baryon-baryon($\varphi B
B'$) coupling, has been discussed by many authors\cite{yukawa} in
the context of the chiral soliton model. In this work, we choose
an operator based on the suggestion of Adkins et al.\cite{ANE} in
relation to the axial current coupling and developed further by
Blotz et al.\cite{BPG}, which has the form\cite{dpp}\cite{ekp}:
\begin{eqnarray}
    \hat O^{(8)}_\varphi = && 3\left[G_0D^{(8)}_{\varphi i}-G_1d_{ibc}
    D^{(8)}_{\varphi b}\hat{S_c}-G_2\frac{1}{\sqrt3}
    D^{(8)}_{\varphi8}\hat{S_i}\right] \nonumber \\
    &&  \times p_{\varphi i},
\end{eqnarray}
where $i=1,2,3$ and $b,c=4,....,7$. The decay amplitude and the
decay width are given by
\begin{equation}
    M_{B \rightarrow B'+ \varphi} = \langle B'|~ \hat{O}_{\varphi}^{(8)} ~
|B
    \rangle, ~ \Gamma_{B \rightarrow B' \phi} = K \cdot \bar{A}^2
\end{equation}
where $\bar{A}^2 = |M|^2/3 p^2$ and $K$ is a kinematic factor :
\begin{equation}
     K = \frac{p^3}{8 ~ \pi ~ M_{B} ~
        M_{B'}}\frac{\overline{M}_{B'}}{\overline{M}_{B}}.
\end{equation}
Here $M_B$'s($m$) are the corresponding masses of the
baryons(mesons), $\overline{M}'s$ are the mean masses of the
multiplet. We take $ \overline{M}_8 = 1154.5 ~MeV , ~~
\overline{M}_{10} = 1436 ~MeV , ~~ \overline{M}_{\overline{10}} =
1726 ~MeV $.

The decay amplitudes of the  baryons can be calculated in a
straightforward way and result in lengthy formulae.  For example,
the amplitudes squared for $\Delta \rightarrow N + \pi$ and $
\Theta^+ \rightarrow N + K$  are given by
\begin{eqnarray}
\bar{A}^2(\Delta \rightarrow N +\pi) &=&
\frac{3}{5}[G_{10}c^{N}_{8}a^{\Delta}_{10}
+\frac{\sqrt{30}}{9}G_{27} c^{N}_{8} a^{\Delta}_{27}
\nonumber\\
&+&\frac{5\sqrt{6}}{18} F_{35} c^{N}_{\bar{10}} a^{\Delta}_{27}
+\frac{1}{3 \sqrt{6}} G'_{27} c^{N}_{27} a^{\Delta}_{10}
\nonumber\\
&+&\frac{\sqrt{5}}{7} G_{27} c^{N}_{27} a^{\Delta}_{27}
+\frac{25}{18}\sqrt{\frac{3}{7}} F_{35}
c^{N}_{27}a^{\Delta}_{35}]^2 \label{delta},\nonumber\\
\bar{A}^2(\Theta^+ \rightarrow N + K)&=&
\frac{3}{5}[G_{\bar{10}}c^{N}_{8}  d^{\Theta ^ +}_{\bar {10}}
+\frac {\sqrt 5}{4} H_{\bar {10}} c^{N}_{\bar {10}} d^{\Theta
^+}_{\bar {10}}\nonumber\\
&-& \frac{7}{4 \sqrt {6}} H'_{27}c^{N}_{27}d^{\Theta ^+}_{\bar
{10}}
]^2, \label{tehta+}\nonumber\\
\end{eqnarray}
where $ G_{10} = G_{0} + \frac{1}{2} G_{1} , G_{27} = G_{0} -
\frac{1}{2} G_{1}, G'_{27} = G_{0} - 2G_{1}, F_{35}  = G_{0} +
G_{1},G_{\bar{10}} = G_{0} - G_{1} - \frac{1}{2} G_{2},
H_{\bar{10}} = G_{0} - \frac{5}{2} G_{1} + \frac{1}{2} G_{2},
H'_{27} = G_{0} + \frac{11}{14} G_{1} + \frac{3}{14} G_{2} $.
Introducing a parameter $\rho$ \cite{praszalowicz2} as $G_1 = \rho
~ G_0 $, we take $G_0$ and $\rho$ as parameters in this
phenomenological analysis. We find a $\rho$ and  $G_0$ that are
consistent with the overall fit to  the experimental values of the
widths of the decuplet. The overall fit is obtained with $G_0 =
17.5$ and $\rho=.5$. The decay width is found to be quite
sensitive to the masses of the particles involved in the decay
process. This is because the kinetic part is very sensitive to the
masses. We calculate the possible range of the calculated widths
by allowing $3\%$ variations of the masses.  As shown in the
parenthesis in Table \ref{tab:table3},  the  kinetic terms $K$ and
therefore the decay widths are changing in a relatively large
range even with $3\%$ variation with masses.  On the other hand by
allowing $\pm 3\%$ variation in masses, reasonably well reproduced
in this model, one can explain the experimental values of $\{10\}$
decay widths within the right range. Now given the set of
parameters determined by the low-lying baryons, one can make the
prediction for the decay widths of exotic $\{\bar{10}\}$ baryons.
In this work we adopt the parametrization for $G_2$ as in
\cite{ekp}, $ G_2 = \left( \frac{9F/D-5}{3F/D +5}\right) (\rho
+2)G_0.$  The estimated decay widths are given in Table
\ref{tab:table4}.

\begin{widetext}

\begin{table*}[h!]
\caption{\label{tab:table3} $\{10\} \rightarrow  \{8\} +\varphi$}
\begin{ruledtabular}
\begin{tabular}{   ccccc  }
Decay  &  K\footnote{Values in the parenthesis are obtained with $\pm 3\%$
mass variations.}   &
   $ \bar{A}^2$ ~  &  $\Gamma$\footnotemark[1]  &  $\Gamma^{Exp.}$  \\
\tableline

&\\

$\Delta \rightarrow N + \pi  $
& 0.33 (0.13$\sim$0.64)~ &  367 ~& 121 (46 $\sim$ 233) ~ & 115 $\sim$ 125 \\

$ \Sigma_{10} \rightarrow \Lambda + \pi $
&   0.17 (0.04$\sim$0.44) ~  &   177 ~ &   31 (8 $\sim$ 79) ~  & 34.7\\

$ \Sigma_{10}  \rightarrow  \Sigma + \pi $
&   0.001 ( $<$ 0.18)   &   43  ~ &   0.70 ( $<$ 7.9)    & 4.73\\

$ \Xi_{10} \rightarrow \Xi + \pi $
&   0.01 ( $<$ 0.22)  &   135  ~ &   1.2 ( $<$ 30)    & 9.9\\
\end{tabular}
\end{ruledtabular}
\end{table*}

\begin{table*}[h!]
\caption{\label{tab:table4} $\{\bar{10}\} \rightarrow \{8\} +
\varphi$}
\begin{ruledtabular}
\begin{tabular}{   cccccc  }
Decay  &  K\footnote{Values in the parenthesis are obtained with $\pm 3\%$
mass variations.}
& ~ $ \bar{A}^2_{(best fit)}$ ~ &  $ \bar{A}^2_{(G_{1}=0, G_{2}=0)}$
&   $ \bar{A}^2_{(without\{27\})}$ & $\Gamma$\footnotemark[1] \\
\tableline

&\\

$ \Theta^+ \rightarrow N + K  $
  & ~  0.52 (0.15 $\sim$ 1.04) ~  &  7.60  &  165  &  29.00  &  4 (1.2
$\sim$ 7.9) \\

$ \Xi_{\bar{10}} \rightarrow \Xi + \pi  $
  & ~  0.66 (0.42 $\sim$ 1.3) ~  &    39  &  125  &  16   & ~  26 (16 $\sim$
50)~ \\

$ \Xi_{\bar{10}} \rightarrow \Sigma + K  $
  & ~  0.23 (0.06 $\sim$ 0.86)  ~  &   17  &  41  &  17   &   4 (0.97 $\sim$
14)\\

$ N_{\bar{10}} \rightarrow N + \pi  $
  & ~  3.3 (2.5 $\sim$ 4.2) ~  &   0.61  &  15.2  &  14  &   2 (1.5 $\sim$
2.6)\\

$ N_{\bar{10}} \rightarrow N + \eta  $
  & ~  1.1 (0.56 $\sim$ 1.8) ~  &   6.6  &  24  &  14  &   7.3 (3.6 $\sim$
12)\\

$ N_{\bar{10}} \rightarrow \Lambda + K  $
  & ~  0.28 (0.01 $\sim$ 0.67) ~ &   3.6  &  31  &  0.7  &   1.0 (0.03
$\sim$ 2.40)\\

$ N_{\bar{10}} \rightarrow \Sigma + K  $
  &   - ( $<$ 0.27) &   0.40  &  45  &  2.3  &   - ( $<$ 0.11)\\

$ \Sigma_{\bar{10}} \rightarrow N + K  $
  &   2.4 (1.6 $\sim$ 3.3)   &   0.50  &  22  &  0.02  &   1.2 (0.81 $\sim$
1.7)\\

$ \Sigma_{\bar{10}} \rightarrow \Sigma + \pi  $
  &   1.3 (1.0 $\sim$ 2.2) &   1.3  &  13  &  12  &   1.7 (1.3 $\sim$ 2.9)\\

$ \Sigma_{\bar{10}} \rightarrow \Sigma + \eta  $
  &   0.06 ( $<$ 0.57)   &   18  &  46  &  34  &   1.0 ( $<$ 10)\\

$ \Sigma_{\bar{10}} \rightarrow \Lambda + \eta  $
  &   0.57 (0.13 $\sim$ 1.1)  &   1.1  &  22  &  1.1&   0.61 (0.14 $\sim$
1.2)\\

$ \Sigma_{\bar{10}} \rightarrow \Xi + K  $
  &   - ( $<$ 0.20)  &   33  &  70  &  91  &   - ( $<$ 6.60)\\

\end{tabular}
\end{ruledtabular}
\end{table*}

\end{widetext}

Compared to the decuplet, the decay amplitudes
for the antidecuplet are found to be much  smaller by an order of
magnitude whereas the kinetic terms are comparable to each other.
It has been understood that this is mainly due to the destructive
interference between the operators\cite{dpp}. In the fourth
column, the amplitudes with $G_1=G_2=0$ are shown, which clearly
shows that the effect of interferences are substantially large. To
see the effect of representation mixing particularly with $\{27\}$
, the results without $\{27\}$ mixing are shown in the fifth
column. The overall tendency is that the non-vanishing mixing with
$\{27\}$ reduces the amplitudes \cite{praszalowicz2}.  However for
the processes $\Xi_{\bar{10}} \rightarrow \Xi + \pi$ and  $
\Sigma_{\bar{10}} \rightarrow N + K$, the mixing enhances the
decay amplitudes\cite{praszalowicz2}, whereas $ \Xi_{\bar{10}}
\rightarrow \Sigma + K  $ and $\Sigma_{\bar{10}} \rightarrow
\Lambda + \eta $ are found to be insensitive to $\{ 27\}$ mixing.
The values in parenthesis are those with $\pm 3~\% $ variations of
the baryon masses.  According to our calculated masses, the
process $ N_{\bar{10}} \rightarrow \Sigma + K$  and
$\Sigma_{\bar{10}} \rightarrow \Xi + K $  are beyond the threshold
in the best fit.

In this work, we discussed the effect of  representation mixing
obtained in $SU(3)$ Skyrme model by diagonalizing the
representation dependent part in the hamiltonian resulting from
quantizing the rotational mode.  It is shown that even without the
next-to- leading order(in $N_C$) symmetry breaking terms  the
low-lying baryon masses can be fairly well reproduced by allowing
the mixing with higher representation. One of the major
differences in the mass results obtained  in the exact
diagonalization method compared to the first order
estimation\cite{dpp} is that there is a deviation from the equal
spacing rule with hypercharge in the $\{ \bar{10} \}$ multiplet .
It is due to the nonnegligible mixing with other
representations\cite{praszalowicz1}. It is also observed that the
mixing effect is not negligible in the decay widths. The effect of
mixing with $\{ 27 \}$ is found to be particularly large. These
results are consistent with the second-order perturbation scheme,
where higher order corrections are  found to be relatively large
\cite{praszalowicz2}.  Although  the decay-width estimations in
this work are based on a specific form of the decay
operator\cite{ANE}\cite{BPG},  the observation that the results of
the exact diagonalization method and the second-order perturbation
scheme are consistent with each other demonstrates that the higher
order corrections beyond the second order might not be important
in numerical estimations. However, it should be noted that the
exact diagonalization can make more sense only when the
hamiltonian to be diagonalized is as complete as possible at least
for the symmetry breaking part.

We would like to thank Mannque Rho for useful discussions and the anonymous
referee for valuable suggestions. We benefited also from the discussions
with Hyun-Chul Kim, Vladmir Kopeliovich, Ghill-Seok Yang and
  Kwang-Yun Youm . This work was
supported by the research fund of Hanyang University (HY-2003-1).

\end{document}